\documentclass[aps,pre,twocolumn,groupedaddress,showpacs]{revtex4}

\usepackage{graphicx}

\begin{document}

\title{The structure of small lead clusters}

\author{Jonathan P.~K.~Doye}
\email[]{jpkd1@cam.ac.uk}
\homepage[]{http://www-wales.ch.cam.ac.uk/~jon}
\affiliation{University Chemical Laboratory, Lensfield Road, Cambridge CB2 1EW, United Kingdom} 
\author{Shaun C. Hendy}
\affiliation{Applied Mathematics, Industrial Research Limited, Lower Hutt, New Zealand}

\date{\today}

\begin{abstract}
We have located the global minimum for all lead clusters with up to 160 atoms using 
a glue potential to model the interatomic interactions. The lowest-energy structures
are not face-centred cubic as suggested previously. Rather, for $N<40$ the majority of structures
are decahedral or hexagonal close-packed, and beyond this size the structures do not correspond to any
of the structural forms commonly found in clusters.  However, these latter clusters
are not simply disordered. High symmetry, magic number clusters are still present, the
most prominent of which is the 148-atom $D_{3d}$ hexagonal barrel.
We relate these structural preferences back to the form of the interactions.
\end{abstract}


\maketitle

\section{\label{sect:intro}Introduction}

The structure of a cluster is one of its primary properties and 
one which has been intensely studied, experimentally and 
theoretically \cite{Johnston02,Alonso00}. 
However, there is still much to be learnt about the fundamentals of 
cluster structure and the possible structures that can be formed. 
For atomic clusters with pair interactions, it is relatively well-understood
how the form of the potential determines the observed structure. For example,
the effects of the width of the potential well \cite{Doye95c,Doye97d} 
and oscillations in the potential \cite{Doye01a,Doye01d,Doye02d} 
have been systematically studied.
However, for the systems that are of most interest, the interatomic
interactions are usually much more complex.
In particular, metal clusters, which are of great technological
relevance \cite{Argo02}, have a strong many-body character to their
bonding.

This presents a number of challenges to our understanding of cluster
structure in metals. First, there is the possibility that new types of 
structure could emerge as a result of many-body effects. 
Although, the structural types observed for pair 
potentials are also frequently observed for metals, e.g.\ the competition
between icosahedral, decahedral and close-packed structures is also 
common for metals \cite{Cleveland91}, 
there are an increasing number of intriguing exceptions.
One seemingly common feature for small metal clusters is to exhibit 
structures with no discernible order 
\cite{Garzon98,Michaelian99,Garzon00,Taneda01,Michaelian02}. 
However, it might be that the disorder is a result of new structural
principles that cannot be fully satisfied at the sizes considered (hence the disorder),
but which could lead to novel high symmetry structures at certain 
magic sizes \cite{Soler01}. Indeed, there are general grounds to expect high symmetry 
structures to emerge irrespective of the potential \cite{Wales98}.
Studies that just reoptimize known cluster structures will of course
miss such new features, and so it is important that efficient global
optimization algorithms are used. 

Secondly, the many-body character makes it increasingly difficult
to relate the observed structure back to the interactions, even
when the assumed form for the many-body potential is relatively simple. 
There has been some interesting progress recently in this area, namely into
the causes of the disordered structures \cite{Soler00}, and the effect of the 
range of the attraction and repulsion on the competition
between icosahedral, decahedral and close-packed clusters \cite{Baletto02b}, 
but there is much still to be discovered. This task is particularly 
important because of the difficulty in producing good empirical metal potentials
(it is not feasible to study the sizes in which we are interested in any
other way). One needs this kind of physical insight to understand the strengths 
and deficiencies of a potential and how it could be improved. It would
also help one to discriminate between different potentials that purport to 
model the same material but give rise to different structures.

Lead clusters illustrate some of these challenges.
The first theoretical study on large clusters by Lim {\it et al.\/} 
using a glue potential seemed to indicate that 
the most stable clusters at relatively small sizes (from at least $N\sim 55$) 
are face-centred cubic (fcc) \cite{Lim}, the preferred bulk structure \cite{Young91}. 
This conclusion was based on a comparison of the energies of 
a series of Mackay icosahedra and fcc cuboctahedra. 
It is quite unusual to see bulk structures already being favoured at such small sizes,
but this finding was rationalized on the basis of the particularly small value of $\gamma$, 
the ratio of the surface energies of the $\{100\}$ and $\{111\}$ faces \cite{Lim}. 
Usually, the Mackay icosahedra have an energetic advantage at small sizes, because exclusively 
having $\{111\}$ facets gives them an appreciably lower surface energy.
Furthermore, these results were not inconsistent with the experiments in the literature at that
time, which were mass spectroscopic studies on very small lead clusters \cite{Muhlbach82,LaiHing87},
and electron diffraction experiments on very large clusters, which were identified as fcc, but
with possibly some vestiges of amorphous structure \cite{Yokozeki78}.

This basic picture though has recently been challenged by 
both experimental \cite{Hyslop01} and theoretical \cite{Hendy01} results.
Electron diffraction of clusters from 3 to 7$\,$nm indicates that the largest 
clusters are dominated by decahedra, but for the smaller clusters it was not 
possible to obtain an adequate fit to the diffraction pattern, suggesting that
alternative structural models need to be considered \cite{Hyslop01}.

Simulations of the melting and freezing of large lead clusters (modelled 
by the glue potential used by Lim {\it et al.\/} \cite{Lim}) unexpectedly revealed 
that for a certain size range ($600<N<4000$, at least) fcc structures are 
not lowest in energy \cite{Hendy01}. 
Instead, a new type of icosahedral structure, which is more stable than 
the fcc structures, spontaneously formed both on freezing 
and on heating at temperatures just below that for melting.
Similar structures had been previously seen in some simulations of large lead
clusters but it was not recognised that they could be lowest in energy \cite{Lim94}.
They resemble anti-Mackay icosahedra \cite{Doye97b}, which have a Mackay icosahedral core 
but with most of the outer layer in `hexagonal close-packed' (hcp) surface sites rather than 
the `fcc' sites that would continue the packing in the Mackay icosahedra.

These results naturally raise intriguing questions about the structures
of small lead clusters.
What alternative structural models might be needed to 
understand the experimental results? 
Do lead clusters, as modelled by the glue potential, really favour fcc structures
at small sizes? Lim {\em et al.} clearly showed that other standard forms were not lower in
energy, so the structures would have to be somewhat unusual.
Interestingly, a simple analysis using macroscopic properties as inputs suggested that 
lead would be a particular likely candidate for disordered clusters to be low in energy \cite{Soler00}.

Here, we address some of these issues by performing global optimization for lead
clusters with up to 160 atoms. We pay particular attention to 
characterizing the structures of these clusters and to 
understanding why the potential favours the lowest-energy structures.

\section{\label{sect:methods}Methods}

To model the lead clusters we use a glue potential \cite{Ercolessi88} of the form
\begin{equation}
E=\sum_{i<j} \phi\left(r_{ij}\right)+\sum_{i} U\left(n_i\right),
\end{equation}
where $\phi(r)$ is a short-ranged pair potential, $U(n)$ is a many-body glue function
and $n_i$ is a ``generalized coordination number'' for atom $i$. $n_i$ is defined as 
\begin{equation}
n_i=\sum_j \rho\left(r_{ij}\right),
\end{equation}
where $\rho(r)$ is an ``atomic density'' function.
These three functions have been fitted for lead using a variety of bulk
and surface properties \cite{Lim}. 
The use of surface energies is particularly important for the application 
of this potential to model clusters. 
As well as clusters, this potential has been also used to model the surface reconstructions 
and pre-melting of low-index lead surfaces \cite{Toh94,Landa95}, and 
lead nanowires \cite{Gulseren95,Gulseren98}.

The choice of this empirical potential is motivated by the need for computational efficiency
in order that global optimization is feasible for the sizes we consider here, and by our
intention to compare with previous results.
The use of {\em ab initio} electronic structure methods for lead is prohibitively
expensive, especially as relativistic effects would need to be included to obtain 
reasonable results \cite{Young91}. 
For example, sophisticated density functional calculations are unable to reproduce the 
experimental surface energies and anisotropies \cite{Vitos98,Feibelmann_both}.

\begin{figure}
\includegraphics[width=8.4cm]{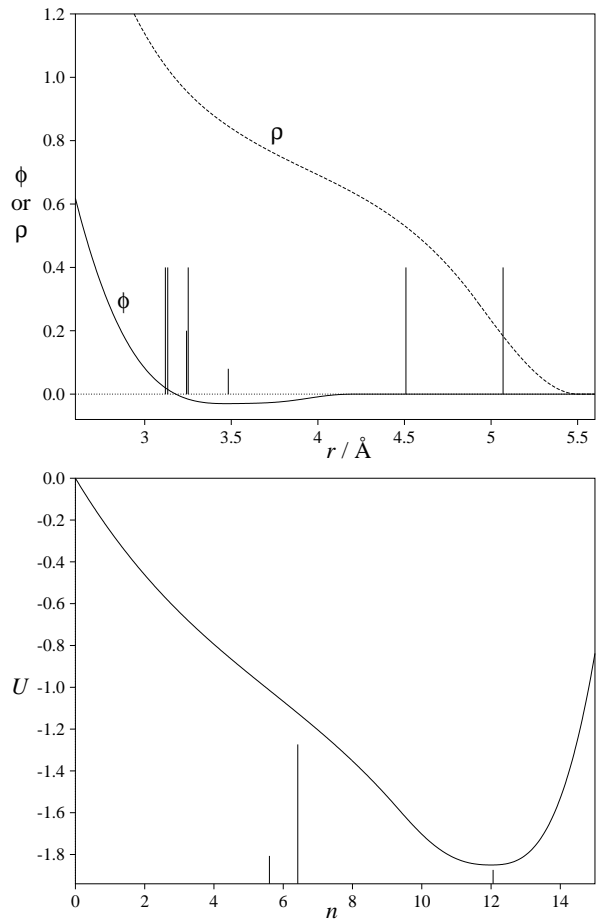}
\caption{\label{fig:potential}The three functions that make up the lead glue potential:
(a) $\phi(r)$, $\rho(r)$ and (b) $U(n)$. The pair distances and $n_i$ values in the 
13-atom decahedron are also plotted as impulses, with the heights proportional to the
number that take that value.}
\end{figure}

The functions $U(n)$, $\phi(r)$ and $\rho(r)$ are displayed in Figure \ref{fig:potential}.
The pair potential has a very shallow well and so most of the binding energy
comes from the glue term. The glue term has been chosen to have its 
minimum at $n$=12, consistent with the designation of $n$ as an effective
coordination number. The form of $\rho(r)$ is particularly significant.
As $\rho(r)$ decreases relatively slowly with increasing $r$ beyond the minimum in the pair potential,
next-nearest neighbours make a significant contribution to $n$. Therefore, 
the difference in surface energies between the $\{111\}$ and $\{100\}$ 
faces is small because, although an atom on a $\{111\}$ face has fewer nearest
neighbours, it has more next-nearest neigbours \cite{Lim}.
However, $\rho(r)$ then decreases relatively rapidly to zero at the cutoff at $r=5.503$\AA,
which typically occurs between the second and third neighbour shells.

For a pair potential the pair distances are the most important quantities.
For example, the lowest-energy structure of a cluster involves a balance
between maximization of the number of nearest neighbours, whilst minimizing 
the strain energy that results from nearest-neighbour pair distances deviating
from the equilibrium pair value, $r_{\rm eq}$ \cite{Doye95c}. 
However, for a glue potential, such as the current one, where the main contribution
to the energy is from the glue function, the most important quantities are the $n_i$.
Indeed, one of the key factors in generating a low-energy structure is to have
the $n_i$ values as close as possible to $n_{\rm eq}$, the value of 
$n$ at the minimum of $U$.
This can potentially lead to different ordering principles than for pair potentials.
Only structures that have their nearest-neighbour pair distances close to $r_{\rm eq}$ are
generally competitive for pair potentials. However, this constraint is relaxed for glue potentials,
and particularly when, as in the current case, $\rho(r)$ initially falls off weakly with $r$.

For the atoms on the surface of a cluster $n_i<n_{\rm eq}$. 
Therefore, there will be a driving force for contraction of the surface to make the pair 
distances for the surface atoms smaller and hence their $n_i$ larger. 
At equilibrium the surface contraction will be balanced by the increase in energy due 
to the resulting compression of the cluster core. 

These considerations represent a particular problem for some of the usual forms 
for atomic clusters, such as the Mackay icosahedra and to a lesser extent decahedra. 
The inherent strain in these clusters results in pair distances between 
surface atoms that are longer than $r_{\rm eq}$, and so the compression needed to 
increase $n_i$ for the surface atoms is particularly large. 
Therefore, these traditional structural forms are expected to become increasingly disfavoured 
by potentials for which the pair separation depends strongly on coordination 
number \cite{Soler00}.
Instead, novel forms that are able to obtain large $n_i$ values for the surface atoms,
whilst not having too large an energetic penalty for compression of the cluster interior,
could potentially be lowest in energy.

The global optimization of the lead clusters was performed using the basin-hopping \cite{WalesD97,WalesS99}
(or Monte Carlo minimization \cite{Li87a}) approach. This method has proved particularly
successful in locating putative global minima for a wide variety of cluster systems \cite{Web}.
The optimization task becomes rapidly more difficult with increasing $N$ (e.g.\ the number of minima 
on the potential energy surface is thought to scale exponentially with $N$ \cite{Tsai93a,Still99,Doye02a}) 
and so, of course, the possibility that we have not been able to obtain the true global minimum increases.
However, the structural principles and trends are clear from our results.

\section{\label{sect:results}Global minima for $N\le 160$}

\begin{figure}
\includegraphics[width=8.4cm]{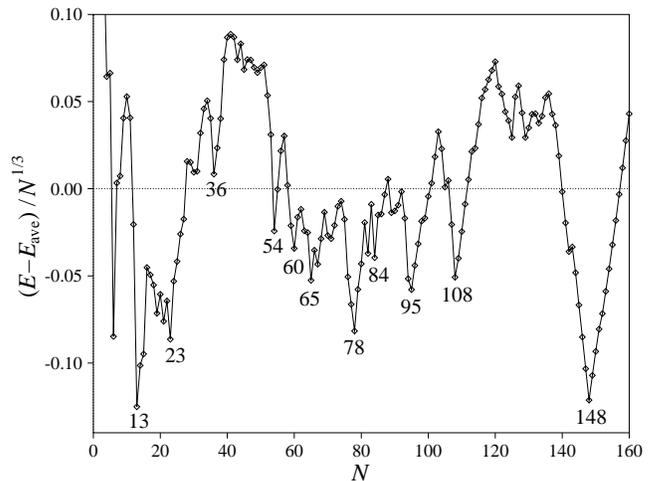}
\caption{\label{fig:Egmin}Energies of the putative global minima relative to $E_{\rm ave}$, 
a four-parameter fit to these energies. $E_{\rm ave}=-2.0251 N+1.6608 N^{2/3}+1.3662 N^{1/3}-0.8634$}
\end{figure}

\begin{figure*}
\includegraphics[width=16cm]{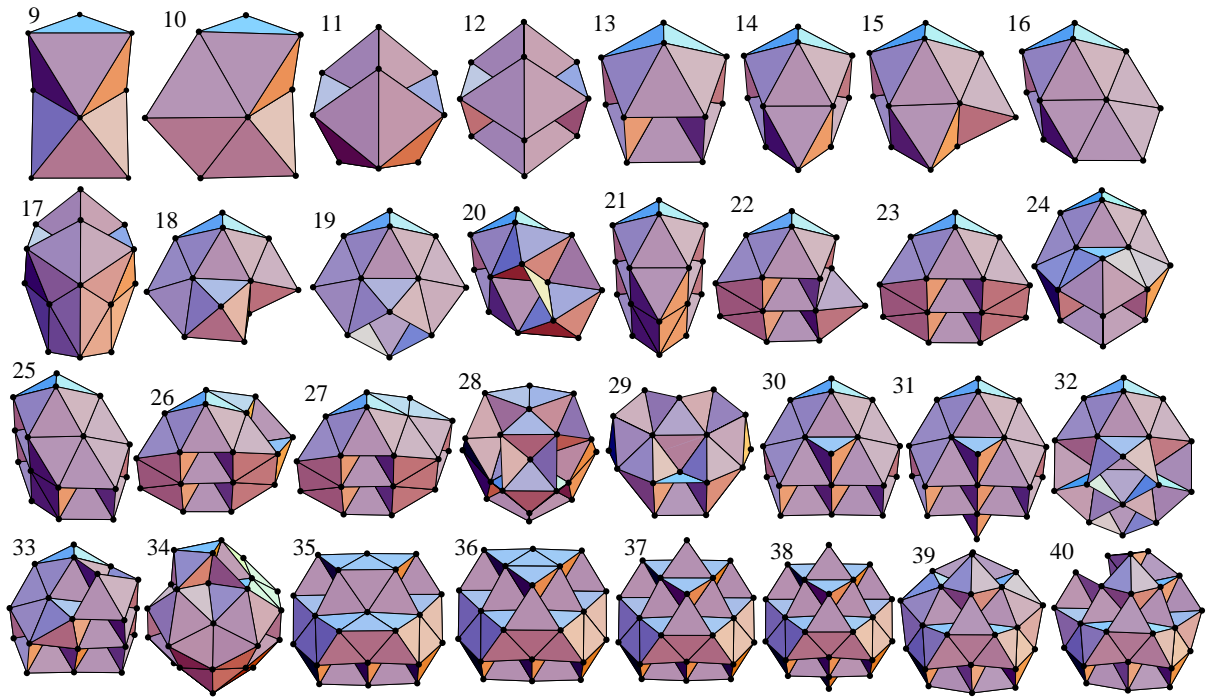}
\caption{\label{fig:Pbsmall} The global minima for $N\le 40$.
Each cluster is labelled by the value of $N$.}
\end{figure*}

\begin{figure*}
\includegraphics[width=16cm]{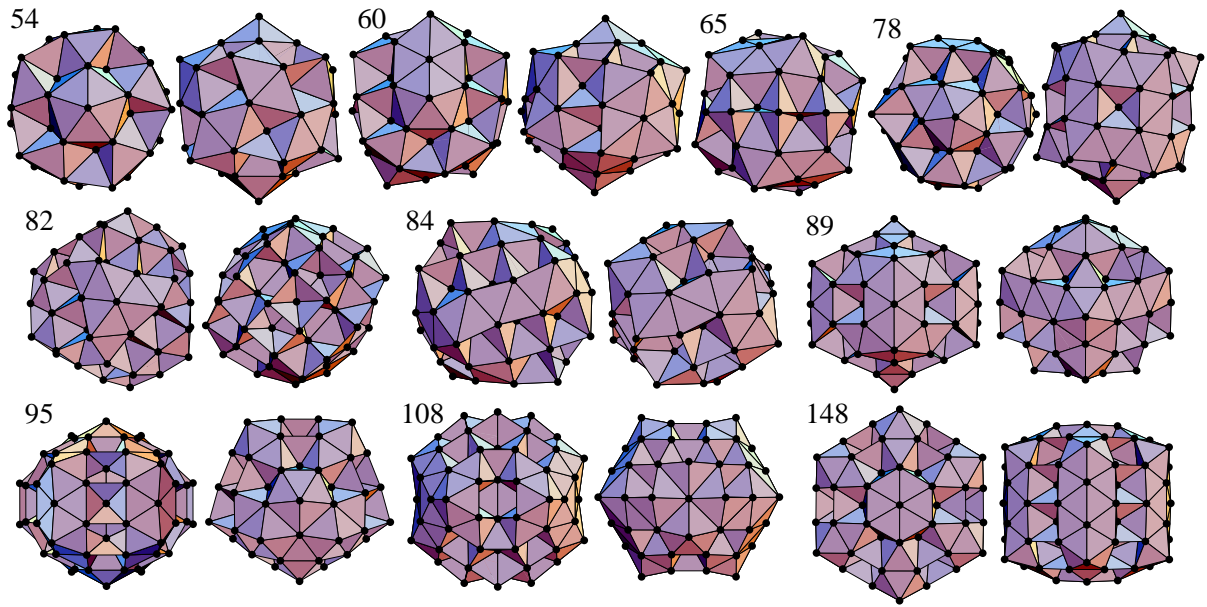}
\caption{\label{fig:Pbbig}A selection of particularly stable global minima for $N>40$.
Each cluster is labelled by the value of $N$. For most of the clusters two perpendicular
views of the structure are given.}
\end{figure*}

The energies and point groups for the putative global minima are given in Table \ref{table:gmin}.
Point files will be made available online at the Cambridge Cluster Database \cite{Web}.
The energies of the global minima are represented in Figure \ref{fig:Egmin} in such a way 
that makes particularly stable clusters stand out. All clusters in the range $9\le N\le 40$
are depicted in Figure \ref{fig:Pbsmall} and a selection of particularly stable 
larger clusters in Figure \ref{fig:Pbbig}.

\begin{table*}
\caption{\label{table:gmin}Energies and point groups (PG) of the putative global minima.}
\begin{ruledtabular}
\begin{tabular}{ccccccccccccccc}
$N$ & PG & Energy/eV & & $N$ & PG & Energy/eV & & $N$ & PG & Energy/eV & & $N$ & PG & Energy/eV \\
\hline
   3 &  $D_{3h}$ &    -1.380851 & &   43 &  $C_s$    &   -62.518056 & &   83 &  $C_1$    &  -131.453946 & &  123 &  $C_2$    &  -201.844885 \\ 
   4 &  $T_d$    &    -2.558548 & &   44 &  $C_s$    &   -64.158589 & &   84 &  $D_2$    &  -133.335862 & &  124 &  $C_2$    &  -203.652142 \\ 
   5 &  $D_{3h}$ &    -3.711742 & &   45 &  $C_s$    &   -65.887013 & &   85 &  $C_1$    &  -134.977705 & &  125 &  $C_1$    &  -205.483134 \\ 
   6 &  $O_h$    &    -5.214277 & &   46 &  $C_1$    &   -67.545017 & &   86 &  $C_2$    &  -136.725287 & &  126 &  $C_1$    &  -207.149663 \\ 
   7 &  $D_{5h}$ &    -6.342793 & &   47 &  $C_1$    &   -69.227454 & &   87 &  $C_1$    &  -138.426910 & &  127 &  $C_1$    &  -208.901679 \\ 
   8 &  $C_{2v}$ &    -7.665775 & &   48 &  $C_1$    &   -70.926964 & &   88 &  $C_1$    &  -140.139735 & &  128 &  $C_1$    &  -210.764087 \\ 
   9 &  $D_{3h}$ &    -8.962242 & &   49 &  $C_2$    &   -72.624639 & &   89 &  $C_{2v}$ &  -141.978803 & &  129 &  $C_1$    &  -212.620619 \\ 
  10 &  $C_{2v}$ &   -10.328111 & &   50 &  $C_1$    &   -74.303601 & &   90 &  $C_s$    &  -143.728686 & &  130 &  $C_1$    &  -214.378335 \\ 
  11 &  $C_{3v}$ &   -11.771970 & &   51 &  $C_1$    &   -75.989247 & &   91 &  $C_{2v}$ &  -145.469445 & &  131 &  $C_1$    &  -216.125616 \\ 
  12 &  $D_{3h}$ &   -13.351511 & &   52 &  $C_1$    &   -77.748769 & &   92 &  $C_1$    &  -147.190877 & &  132 &  $C_1$    &  -217.911017 \\ 
  13 &  $D_{5h}$ &   -15.060197 & &   53 &  $C_1$    &   -79.529727 & &   93 &  $C_1$    &  -149.017327 & &  133 &  $C_1$    &  -219.726303 \\ 
  14 &  $C_{2v}$ &   -16.488673 & &   54 &  $S_{10}$ &   -81.438379 & &   94 &  $C_{2v}$ &  -150.933479 & &  134 &  $C_1$    &  -221.493988 \\ 
  15 &  $C_{2v}$ &   -17.971698 & &   55 &  $C_1$    &   -83.050627 & &   95 &  $C_{2v}$ &  -152.722173 & &  135 &  $C_1$    &  -223.228659 \\ 
  16 &  $C_{2v}$ &   -19.359118 & &   56 &  $C_1$    &   -84.670414 & &   96 &  $C_s$    &  -154.419817 & &  136 &  $C_1$    &  -225.006412 \\ 
  17 &  $C_{3v}$ &   -20.892141 & &   57 &  $C_1$    &   -86.343451 & &   97 &  $C_1$    &  -156.124859 & &  137 &  $C_1$    &  -226.856548 \\ 
  18 &  $C_1$    &   -22.441282 & &   58 &  $C_1$    &   -88.160853 & &   98 &  $C_1$    &  -157.827739 & &  138 &  $C_2$    &  -228.680493 \\ 
  19 &  $C_{2v}$ &   -24.029140 & &   59 &  $C_1$    &   -89.961499 & &   99 &  $C_1$    &  -159.582628 & &  139 &  $C_1$    &  -230.562421 \\ 
  20 &  $C_{2v}$ &   -25.554526 & &   60 &  $C_1$    &   -91.725507 & &  100 &  $C_1$    &  -161.288978 & &  140 &  $C_1$    &  -232.461447 \\ 
  21 &  $C_{2v}$ &   -27.160557 & &   61 &  $C_1$    &   -93.369916 & &  101 &  $C_1$    &  -163.018692 & &  141 &  $C_1$    &  -234.347193 \\ 
  22 &  $C_1$    &   -28.700367 & &   62 &  $C_1$    &   -95.068322 & &  102 &  $C_1$    &  -164.713475 & &  142 &  $C_1$    &  -236.226662 \\ 
  23 &  $C_{2v}$ &   -30.342369 & &   63 &  $C_1$    &   -96.835577 & &  103 &  $C_1$    &  -166.412388 & &  143 &  $C_1$    &  -238.007461 \\ 
  24 &  $D_{2h}$ &   -31.834411 & &   64 &  $C_2$    &   -98.559680 & &  104 &  $C_1$    &  -168.225729 & &  144 &  $C_2$    &  -239.879381 \\ 
  25 &  $C_{2v}$ &   -33.394629 & &   65 &  $C_2$    &  -100.391525 & &  105 &  $C_1$    &  -170.097176 & &  145 &  $C_s$    &  -241.772812 \\ 
  26 &  $C_s$    &   -34.947504 & &   66 &  $S_4$  &  -102.045664 & &  106 &  $C_1$    &  -171.848423 & &  146 &  $C_2$    &  -243.665845 \\ 
  27 &  $C_{2v}$ &   -36.526823 & &   67 &  $C_1$    &  -103.803493 & &  107 &  $C_s$    &  -173.738048 & &  147 &  $C_s$    &  -245.558654 \\ 
  28 &  $C_1$    &   -38.036722 & &   68 &  $C_1$    &  -105.470725 & &  108 &  $C_{2v}$ &  -175.653403 & &  148 &  $D_{3d}$ &  -247.451751 \\ 
  29 &  $C_s$    &   -39.653184 & &   69 &  $C_1$    &  -107.136603 & &  109 &  $C_{2v}$ &  -177.373836 & &  149 &  $C_1$    &  -249.175286 \\ 
  30 &  $C_s$    &   -41.291166 & &   70 &  $C_1$    &  -108.921031 & &  110 &  $C_s$    &  -179.073251 & &  150 &  $C_s$    &  -250.900806 \\ 
  31 &  $C_{2v}$ &   -42.914946 & &   71 &  $C_2$    &  -110.659704 & &  111 &  $C_{2v}$ &  -180.771206 & &  151 &  $C_1$    &  -252.632151 \\ 
  32 &  $C_{2v}$ &   -44.475555 & &   72 &  $C_1$    &  -112.360587 & &  112 &  $C_s$    &  -182.477331 & &  152 &  $C_2$    &  -254.384256 \\ 
  33 &  $C_1$    &   -46.064990 & &   73 &  $C_1$    &  -114.048803 & &  113 &  $C_s$    &  -184.174089 & &  153 &  $C_1$    &  -256.115829 \\ 
  34 &  $C_1$    &   -47.687973 & &   74 &  $C_1$    &  -115.771829 & &  114 &  $C_2$    &  -185.939956 & &  154 &  $C_2$    &  -257.846941 \\ 
  35 &  $C_s$    &   -49.363201 & &   75 &  $C_1$    &  -117.552036 & &  115 &  $C_1$    &  -187.649103 & &  155 &  $C_1$    &  -259.573483 \\ 
  36 &  $D_{3d}$ &   -51.115236 & &   76 &  $C_2$    &  -119.429639 & &  116 &  $C_1$    &  -189.351898 & &  156 &  $C_2$    &  -261.299215 \\ 
  37 &  $C_{3v}$ &   -52.716480 & &   77 &  $C_1$    &  -121.236712 & &  117 &  $C_1$    &  -191.105099 & &  157 &  $C_1$    &  -263.019401 \\ 
  38 &  $D_{3d}$ &   -54.314542 & &   78 &  $C_2$    &  -123.043586 & &  118 &  $C_1$    &  -192.855140 & &  158 &  $C_2$    &  -264.738883 \\ 
  39 &  $C_1$    &   -55.857448 & &   79 &  $C_2$    &  -124.684422 & &  119 &  $C_1$    &  -194.607282 & &  159 &  $C_1$    &  -266.456122 \\ 
  40 &  $C_2$    &   -57.474215 & &   80 &  $C_1$    &  -126.364903 & &  120 &  $C_1$    &  -196.361528 & &  160 &  $C_1$    &  -268.174733 \\ 
  41 &  $C_s$    &   -59.130894 & &   81 &  $C_1$    &  -128.007123 & &  121 &  $C_1$    &  -198.211397 \\ 
  42 &  $C_s$    &   -60.803249 & &   82 &  $D_3$  &  -129.829950 & &  122 &  $C_1$    &  -200.012938 \\ 
\end{tabular}
\end{ruledtabular}
\end{table*}

First, we will look at the global minima for $N\le 40$ in detail, before surveying more 
briefly the results for larger clusters.
For $N\le 8$ the clusters exhibit the same structures as typically seen for pair potentials.
However, Pb$_9$ has a somewhat unusual form that can be described as two 
face-sharing octahedra, and so, as with Pb$_{10}$, is the beginning of an hcp cluster. Then
for $N$=11 and 12 more open structures with three-fold axes of symmetry are preferred.

Most of the global minima for $13\le N \le 33$ are decahedral in origin. However, the growth
sequence is not straightforward. The decahedra are generally asymmetric with the quasi-fivefold
axis not passing through the centre of mass. So, although the growth sequence begins by adding atoms
around the equator of the 13-atom Ino decahedron \cite{Ino}, before this shell is completed, asymmetric decahedra 
with a longer quasi-fivefold axis become lower in energy, starting at $N$=21. Furthermore, sometimes
part of the structure is distorted away from the ideal decahedral positions, 
e.g.\ at $N$=18, 19, 26 and 32. There are also structures with two interpenetrating (Pb$_{20}$) and 
face-sharing (Pb$_{24}$) 13-atom decahedra, the latter with two additional shared capping atoms.
It is noticeable that the decahedral global minima generally  have a significant proportion of surface 
atoms in $\{100\}$ type environments. 
This feature reflects the small energy difference between fcc $\{111\}$ and $\{100\}$
faces noted earlier. For materials that more strongly favour $\{111\}$ faces, the most stable
decahedral form is usually a Marks decahedron \cite{Marks84}, because this structure maximizes the 
proportion of $\{111\}$ faces, whilst retaining a relatively spherical shape. 
However, for lead the most stable decahedral clusters occur at 
$N$=13 and 23 (Figure \ref{fig:Egmin}).

The other set of ordered global minima found for $N\le 40$ are the hcp clusters at $N$=35-38. 
Again, these structures are somewhat unexpected, particularly as the fcc truncated octahedron
is possible at $N$=38, but this is further evidence of a preference for structures
with a significant proportion of $\{100\}$-like faces.

Of the other global minima for $N\le 40$, Pb$_{17}$ is related to the 11-atom global minimum, 
but it is hard to discern any overall order for those at $N$=28, 29 and 34. 
Pb$_{39}$ and Pb$_{40}$ are somewhat related to the preceding hcp structures, as is clear from
the viewpoint chosen for Figure \ref{fig:Pbsmall}, but again there is little order apparent
on the other side of the cluster.

Beyond $N$=38 none of the global minima that we have located can be assigned to any of the 
usual structural forms. However, it would be too simplistic just to characterize the clusters 
as disordered. From Table \ref{table:gmin} we can see that high symmetry structures are still 
present. Furthermore, if the clusters were just disordered one would expect cluster properties 
to evolve fairly smoothly with size. However, it is clear from Figure \ref{fig:Egmin} that
there are ``magic number'' clusters that are particularly low in energy. 
Unsurprisingly, these magic numbers often correspond to the high symmetry clusters.

Although most of the clusters in this size range have no discernible {\it overall order}, 
there are common local surface motifs that are repeatedly visible. However, only at a relatively few 
sizes can these local preferences 
be assembled into a structure that has clear overall order. 

As for the smaller clusters the surface structures of the global minima 
reflect the particularly low value of $\gamma$. 
However, this does not lead to structures with large $\{100\}$ faces, but rather
to many surface atoms with $\{100\}$-like environments. The surfaces are typically covered
with a patchwork of squares and triangles. So on the flat regions of the surface it is common to 
see atoms surrounded by three triangles and two squares (there are two ways this can be achieved),
rather than the six triangles or four squares, that are typical of $\{111\}$ and $\{100\}$ surfaces,
respectively.  

Pb$_{54}$ is somewhat related to the Mackay icosahedron. 
It has an uncentred 13-atom icosahedron at its centre and a clear five-fold axis of symmetry. 
Along this axis it looks similar to the the $D_{5h}$ structure that was found by Wolf and Landman 
as a low-energy isomer of the 55-atom Lennard-Jones clusters \cite{Wolf98}, 
and which is related to the icosahedron by a single rearrangement in which 
the structure is twisted around a five-fold axis. However, 
there is a canted arrangement of squares and triangles around the equator of the cluster.

The axial configuration of Pb$_{54}$ seems to be quite a common motif, and 
similar patterns can be seen in one of the chosen views for $N$=60, 78, 95 and 148, 
the last based on a six-fold rather than a five-fold symmetric version of the pattern, thus making the
top surface flat, rather than pyramidal. 
As the size of these clusters increases the pattern is, of course, extended outwards. 
The resulting motif is clearest for the highly symmetric, 148-atom global minimum.

Pb$_{148}$ is the most prominent magic number in this size range (Figure \ref{fig:Egmin}). 
In shape, the cluster is a hexagonal barrel. Although the outer surface has a clear sixfold 
symmetry, this is in fact broken by the octahedron at the centre of the cluster.

\begin{table}
\caption{\label{table:Edecomp}
The contributions to the energy 
for a series of 55-atom structures, namely the global minimum ($C_1$), 
the fcc cuboctahedron ($O_h$), the Ino decahedron ($D_{5h}$) and the Mackay icosahedron ($I_h$).
The structures are denoted by their point group (PG).
$\langle E_i^{\rm bulk}\rangle$ and $\langle E_i^{\rm surf}\rangle$ are the average atomic
energies for atoms in the interior of the cluster and on the surface, respectively.
All the energies are measured in eV.
}
\begin{ruledtabular}
\begin{tabular}{ccccccccc}
 PG & Energy & $E_{\rm pair}$ & $n_{\rm nn}$ & $E_{\rm strain}$ & $E_{\rm glue}$ & 
 $\langle n_i\rangle$ & $\langle E_i^{\rm bulk}\rangle$ & $\langle E_i^{\rm surf}\rangle$ \\
\hline
$C_1$    & -83.051 & -1.796 & 216 & 4.684 & -81.254 & 9.072 & -1.885 & -1.405 \\
$O_h$    & -82.559 & -4.338 & 216 & 2.142 & -78.220 & 8.783 & -1.956 & -1.360 \\
$D_{5h}$ & -82.438 & -4.248 & 219 & 2.322 & -78.190 & 8.800 & -1.933 & -1.365 \\
$I_h$    & -81.295 & -5.944 & 234 & 1.076 & -75.351 & 8.500 & -1.946 & -1.333 \\
\end{tabular}
\end{ruledtabular}
\end{table}

As flagged in the introduction, an important aim of this paper is 
not only to characterize the global minima for this lead potential, 
but to understand how the observed structures relate back to the form 
of the potential.
We start by examining the decahedral 13-atom global minimum, for which the $r_{ij}$ and $n_i$ values
have been included in Figure \ref{fig:potential}. It is noticeable that there is a
significant dispersion of nearest-neighbour distances. In fact the longest distance is 11.7\% longer than
the shortest, which compares to a 2.3\% difference for the same structure when optimized for the
Lennard-Jones potential.
As expected from the discussion in Section \ref{sect:methods}, the structure
distorts to move the $n_i$ values as close to $n_{\rm eq}$ as possible,
rather than keeping the nearest-neighbour distances near to the minimum of the pair potential.
This is achieved by an expansion along the fivefold axis and 
a contraction of the equator of the cluster. This reduces the $n_i$ values for the two vertex atoms
on the fivefold axis, but increases the $n_i$ values for the other ten surface atoms,  while
maintaining the $n_i$ value for the central atom close to $n_{\rm eq}$.
A similar anisotropy of the pair distances has previously been noted by Lim {\it et al.\/}
in their analysis of the lead cuboctahedra \cite{Lim};
there is a greater contraction for the $\{100\}$ faces of the cuboctahedra than the $\{111\}$ faces
because of the enhanced contribution to $n_i$ from next neighbours across the diagonals of the 
squares on the $\{100\}$ faces.

\begin{figure}
\includegraphics[width=8.4cm]{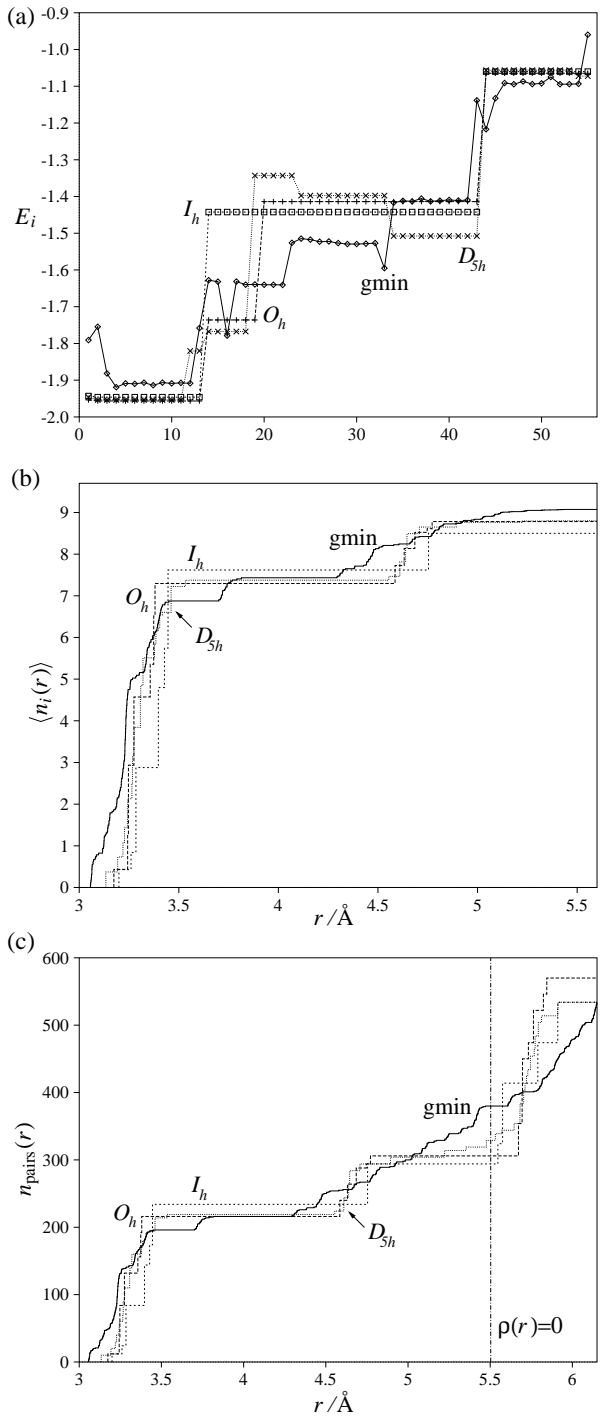}
\caption{\label{fig:analyse}A comparison of the properties of the 55-atom global minimum (gmin)
to the fcc cuboctahedron ($O_h$), the Ino decahedron ($D_{5h}$) and the Mackay icosahedron ($I_h$), 
the same four clusters as in Table \ref{table:Edecomp}.
(a) The atomic energies, $E_i$, for each atom in the cluster.
The atoms have been ranked by their distance from the centre of mass, with atom 1 
being the closest to the centre.
(b) $\langle n_i(r_{ij}<r)\rangle$. (c) $n_{\rm pairs}(r)$.
}
\end{figure}

To understand why novel structural forms are observed for this lead potential, we 
take Pb$_{55}$ as an example and compare the contributions to the energy from 
a series of competing structures (Table \ref{table:Edecomp} and Figure \ref{fig:analyse}).
The global minimum is based on the 54-atom structure illustrated in Figure \ref{fig:Pbbig}
but with an additional surface atom. Also possible at this size are a fcc cuboctahedron, 
an Ino decahedron and a Mackay icosahedron.

In Table \ref{table:Edecomp} we have decomposed the pair energy into two components:
\begin{equation}
E_{\rm pair}=-n_{\rm nn} \epsilon + E_{\rm strain},
\end{equation}
where $n_{\rm nn}$ is the number of nearest neighbours, $\epsilon$ is the depth of the 
pair potential, and $E_{\rm strain}$ is the energetic penalty for distances that deviate from
$r_{\rm eq}$, the distance corresponding to the minimum of the pair potential. 
More formally,
\begin{equation}
E_{\rm strain}=\sum_{i<j,r_{ij}<r_0} \epsilon - \phi(r_{ij}),
\end{equation}
where $r_0$ is a distance criterion that distinguishes nearest from next-nearest neighbours. 
For all the structures we consider, there is a clear separation between these coordination shells.

As expected the pair energy only contributes a small fraction of the total energy. It is also 
noticeable that $E_{\rm strain}$ is of similar magnitude to $E_{\rm pair}$.
This is in marked contrast to what occurs for pair potentials, where minimization of the 
strain energy is a key element of a structure's stability. $E_{\rm strain}$ for the global 
minimum is particularly large. 

Although the pair energy is small in magnitude, it is structure sensitive and so it can still 
determine the relative stabilities of structures when the energies from the glue term, $E_{\rm glue}$,
are similar. For example, the major component of the difference in energy between 
the 55-atom cuboctahedron and decahedron is the greater pair energy of the cuboctahedron.

It is clear from Table \ref{table:Edecomp} that the global minimum's stability is a result of 
its significantly lower glue energy, which is a result of the atoms being able to 
achieve $n_i$ values that are closer to the ideal value, $n_{\rm eq}$.
However, this lower glue energy is partially offset by the higher pair energy 
resulting from the distortion of the pair distances that is necessary to
achieve an increase in $n_i$.

If we look at the atomic contributions to the energy 
(Table \ref{table:Edecomp} and Figure \ref{fig:analyse}(a))
it is clear that the lower energy results from a lower average energy for the
surface atoms (particularly atoms 13--33 in Figure \ref{fig:analyse}(a)),
which outweighs the somewhat less favourable energies for the atoms 
in the interior of the cluster.

It is also interesting to understand what structural features of the
global minimum lead to the larger value of $\langle n_i \rangle$.
We analyse this in Figure \ref{fig:analyse}(b) and (c), first by looking at
the cumulative contribution to $\langle n_i \rangle$ from pairs with distances less than $r$:
\begin{equation}
\langle n_i(r_{ij}<r)\rangle={1\over N} \sum_{i\ne j, r_{ij}<r} \rho(r_{ij}). 
\end{equation}
It is particularly interesting to note that $\langle n_i^<(r)\rangle$ for the global minimum
only becomes largest beyond 4.926\AA. Therefore, although the contribution to 
$\langle n_i \rangle$ from distances beyond this distance is small in magnitude, it
is key in stabilizing the global minimum relative to the more conventional forms.

We can analyse this further by considering $n_{\rm pairs}(r)$, 
the number of pairs of atoms that are separated by less than $r$.
It can be seen from Figure \ref{fig:analyse}(c) that the number of pair distances within the radius of the
cutoff distance for $\rho$ is significantly larger for the global minimum  
than for the competing structures,
which in turn correlates with the larger value of $\langle n_i\rangle$.
However, this is only true because the cutoff is located between the second and third coordination shell.
The cuboctahedron, decahedron and icosahedron all have a relatively narrow distribution of
nearest-neighbour distances, which leads to a clear distinction between the second and third 
coordination shell. By contrast, the global minimum has a much more disperse nearest-neighbour shell and 
hence there is no clear distinction between a second and third neighbour shell. Instead,
there is a steady increase in $n_{\rm pairs}(r)$ beyond the start of the second neighbour shell.

Although the above analysis has been presented for a single cluster, repeating
this procedure for other sizes has confirmed the generality of the conclusions.

\section{\label{sect:conc}Conclusions}

\begin{figure}
\includegraphics[width=8.4cm]{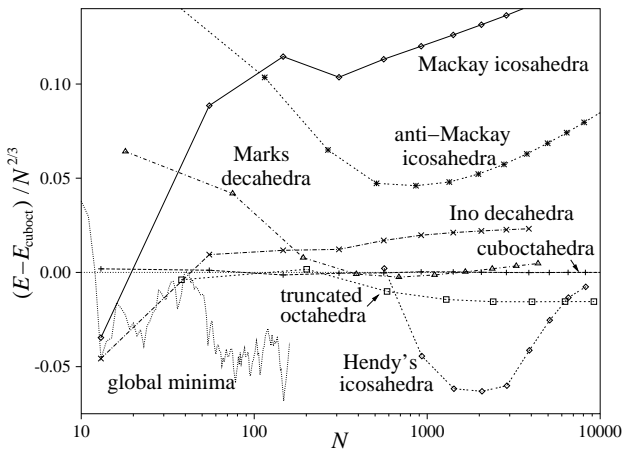}
\caption{\label{fig:E_large} A comparison of the energies of the global minima and 
the new icosahedra discovered by Hendy \cite{Hendy01,Hendy02}, 
to series of high-symmetric structures, which 
include Mackay and anti-Mackay icosahedra, cuboctahedra, truncated octahedra (with regular
hexagonal $\{111\}$ faces), Ino decahedra and Marks decahedra.
The energies are measured with respect to 
$E_{\rm cuboct}=-2.0284 N + 1.7929 N^{2/3} +0.9714 N^{1/3}-0.6342$.
}
\end{figure}

We have shown by the locating the global minima for small lead clusters
interacting with a many-body potential of the glue form that, 
contrary to the original conclusion of Lim {\it et al.\/}, these
clusters do not adopt fcc geometries for $N\le 160$. Instead, they
form a series of novel structures that are a consequence of the many-body
character of the potential. These results naturally lead one to wonder
at what size bulk-like fcc structures will develop. 
To help us answer this question we have plotted in Figure \ref{fig:E_large} 
the energies of the global minima, alongside
those for a number of sequences of high-symmetry structures
and the novel icosahedral forms that Hendy obtained by simulations of freezing \cite{Hendy01}
and by construction \cite{Hendy02}.

The figure confirms that the icosahedra and decahedra are always higher in energy than 
the best fcc structures (except at the smaller sizes considered in the last section) and 
that the fcc truncated octahedra become slightly lower in energy than the 
cuboctahedra \cite{Lim,Hendy01}.
More interestingly, we can clearly see that both the global minima we have found and 
the new icosahedra produced by Hendy \cite{Hendy01,Hendy02} are significantly lower in energy than the best fcc 
clusters. Thus, extrapolations between these two sizes ranges \cite{Hendy_range} suggest that fcc clusters might well
not be the lowest in energy for intermediate sizes, and hence that fcc clusters are not
lowest in energy until at least $N\sim 15\,000$ \cite{Hendy02}.
This is a surprising result, for although it is not uncommon to find small ($N<100$)
metal clusters that do not exhibit any of the usual cluster structures, 
it is unprecedented for this behaviour to persist up to such large sizes.

The results are also relevant to the ongoing issue of disordered metal clusters.
Like recent theoretical results for gold \cite{Garzon98,Michaelian99,Garzon00}, 
cadmium, zinc \cite{Michaelian02} and vanadium \cite{Taneda01}
many of our global minima do not fit with the fcc, hcp, decahedral and icosahedral
structures that are often found for close-packed materials. However, 
to call these lead clusters disordered would be too strong
because, although most of the clusters for $N>40$ have no overall stuctural order,
there are common local structural preferences which at a few sizes result
in highly symmetric ordered structures that are particularly stable.

For pair potentials, clusters tend to retain a lattice structure away from the magic numbers,
because it is unfavourable for the pair distances to deviate significantly from the
equilibrium value. For example, most small Lennard-Jones clusters can be considered to be based upon
Mackay icosahedra, either with an incomplete outer layer or covered by an ordered overlayer \cite{Northby87}.
By contrast, for metal clusters the many-body character of the 
bonding can make it favourable to break the lattice structure away from the magic numbers 
so that the atoms can (in the language of the current potential) 
increase their effective coordination numbers ($n_i$'s). 
A structure with no overall order results. 
Therefore, if one only examines a few cluster sizes the 
presence of particularly stable ordered structures may be missed.

From our analysis of the energetics of the competing structural
forms we have seen that the shape of $\rho(r)$, in particular the shoulder
and the position of the cutoff, is key to the stability of the novel
structures that we find to be lowest in energy. This dependence on
the cutoff is somewhat worrying both because it is a rather long-range feature of the
potential and because its position is not physically motivated, 
but chosen more for computational convenience.
These results illustrate how sensitively cluster structure depends on the potential;
the correct determination of the relative energies of competing clusters is 
a stringent test of any potential \cite{Doye02b}.

Our results also help us to understand the structures exhibited by lead nanowires \cite{Gulseren98}
modelled using the same potential.
G\"ulseren {\it et al.\/} were surprised at the apparent contradiction between
the non-fcc character of their nanowires and the cluster results of 
Lim {\it et al.\/} \cite{Lim}, and so suggested a number of reasons for the
differing structural tendencies. However, our results show that the non-fcc character
is common to both systems.

\begin{acknowledgments}
J.P.K.D is grateful to the Royal Society for the award of a University Research Fellowship.
S.C.H. would like to acknowledge the support of the ISAT 
linkages fund administered by the Royal Society of New Zealand.
\end{acknowledgments}


\begin{thebibliography}{48}
\bibitem[{Hen()}]{Hendy_range}
\bibinfo{note}{Although only one series of structures with the novel
  icosahedral form has been included in Fig.\ \ref{fig:E_large}, many are
  possible, which are characterized by the number of facets that are of a
  certain type \cite{Hendy02}. If the energies of all these series were
  calculated, the size range over which these novel icosahedra are most stable
  would be significantly wider than suggested by Fig.\ \ref{fig:E_large}.}
\end{thebibliography}
\end{document}